\documentstyle[sprocl]{article}

\def\Journal#1#2#3#4{{#1} {\bf #2}, #3 (#4)}

\def\PLB{{\em Phys. Lett.}  B}
\def\PRL{\em Phys. Rev. Lett.}
\def\PRD{{\em Phys. Rev.} D}

\def\be{\begin{equation}}
\def\ee{\end{equation}}
\def\bea{\begin{eqnarray}}
\def\eea{\end{eqnarray}}

\begin{document}

\title{ELECTROWEAK BARYOGENESIS WITHOUT THE PHASE TRANSITION}

\author{MICHAEL JOYCE}

\address{School of Mathematics, Trinity College,
Dublin 2, Ireland.}



\maketitle\abstracts{
Radiation domination at the electroweak epoch is a simplifying 
assumption, but one for which there is no observational basis.
Treating the expansion rate as a variable, I re-examine 
electroweak baryogenesis in various scenarios. At a first
order phase transition the main effect is on the sphaleron bound,
which becomes a lower bound on the expansion rate in any given 
theory. At a second-order or cross-over phase transition, the 
created baryon asymmetry is directly proportional to the 
expansion rate. I sketch an alternative post-inflationary 
cosmology, in which the kinetic energy of a scalar field 
dominates the Universe until shortly before nucleosynthesis, 
and argue that the observed baryon asymmetry could be produced 
in this case even at an analytic cross-over.}
  
\section{Introduction}

A loose analogy between nucleosynthesis and electroweak 
baryogenesis is often drawn in arguing the particular merit
of the latter - that, in the not too distant future,
our experimental knowledge of 
the relevant physics may
be on the same firm basis as our knowledge of the nuclear
and particle physics on which nucleosynthesis rests. 
It has some hope of being a truly testable theory which may 
provide us with some solid probe of the pre-nucleosynthesis 
universe, potentially as compelling as that provided 
by nucleosynthesis of its epoch. 
The analogy is worth pushing a little further.
One of the great successes of nucleosynthesis is how
it has been able 
to constrain the number of effective relativistic degrees of freedom.
This constraint is actually just one on the expansion rate
at freeze-out of the weak interactions, and affects (to a first
approximation) only the abundance of Helium. It is  
a constraint on the cosmology (albeit stated as one on particle physics), 
given this single observational input. In this contribution I 
address the analogous question of electroweak 
baryogenesis: What constraint does the {\it requirement} that the 
observed baryon asymmetry be generated at the electroweak scale place 
on cosmology at that scale? Having answered this question I sketch 
one simple cosmology in which baryon generation could occur at
the electroweak scale in a regime in which it is usually
assumed to be impossible. The results sketched here are from recent
work by myself \cite{mjprd} and in collaboration with T. Prokopec 
\cite{mjtppreprint}, in which the reader can find greater detail
and a fuller set of references.

A trivial but important point to emphasize is that we actually know 
{\it nothing} of cosmology at the electroweak scale: We have
at present no convincing experimental probe of that epoch.
In red-shift the electroweak scale is approximately as far 
from nucleosynthesis as the latter is from the transition to
matter domination, and the simplest backward extrapolation 
is certainly worth calling into question. 
Here I assume as usual that the Universe is
indeed a hot plasma back to temperatures $T > T_{ew} \sim 100$GeV, 
but allow the expansion rate to be a variable $H_{ew}$.
To determine the relics from this epoch we need to know
the expansion rate as a function of temperature around
$T_{ew}$ and take  
\begin{equation}
H=H_{ew} \big(\frac{T}{T_{ew}} \big)^p
\label{tempdep}
\end{equation}
where $p$ is some number, of which the results below
are essentially independent.

\section{EWB at a first order phase transition}

The asymmetry created as envisaged in the standard scenario
as bubble walls sweep through the unbroken phase is essentially
insensitive to a change, even of many orders of magnitude, in
the expansion rate away from its radiation dominated value 
$H_{rad} \sim 10^{-16} T$. The main effect is on the 
sphaleron bound for the preservation of the created asymmetry
which requires that 
\begin{equation}
{\cal D} \equiv -\ln \frac{B_{\rm final}}{B(T_{b})} 
= \int_{t_b}^{\infty} dt
 \Gamma_{sph}(t) 
=H_{b}^{-1}
\int_{0}^{T_{ b}}dT
\frac{\Gamma_{ sph}}{T}
\big(\frac{T_{ b}}{T}\big)^{p} < 1
\label{eq: depletion}
\end{equation}
where $B(T_{b})$ is the baryon asymmetry at the completion of the transition, 
at temperature $T_{b}$ when the expansion rate is $H_b$, and 
$\Gamma_{sph}$ is the appropriately normalized rate of sphaleron processes. 
The latter form of this expression follows from (\ref{tempdep})
which gives 
$t \propto T^{-p}$.
This integral is dominated by temperatures very close to $T_b$
so that the result is essentially independent of $p$. 
The bound can be cast
to a good approximation in the form of an expansion rate dependent 
correction to its familiar form as  
a lower bound on $\frac{\phi_b}{T_b}$
(the ratio of the vev to the temperature at $T_b$):
\begin{equation}
\frac{\phi_b}{T_b} > \left(\frac{\phi_b}{T_b}\right)_{rad} -\frac{0.06}
{\cal{B}}
\ln\frac{H_b}{H_{rad}}
\label{shift}
\end{equation}
where 
$(\frac{\phi_b}{T_b})_{rad}$ is the appropriate critical value 
(typically $1 - 1.2$) in the radiation dominated universe, and
${\cal{B}}\in{[1.5,2.7]}$ is the usual monotonic function of $m_H^2/m_W^2$ 
which appears in relating the sphaleron energy to the vev.
For $m_H \sim 80$GeV, for example, this means a change of $0.08$
in the bound per order of magnitude in the expansion rate. 
In any particular model (\ref{shift}) can alternatively be used to
state the sphaleron bound for preservation of the baryon asymmetry
as a lower bound on the expansion rate of the Universe.

\section { EWB in a homogeneous Universe }

Under this rubric I include any case where the electroweak 
sphaleron processes freeze out at a time when the evolution of the
plasma is well approximated as homogeneous. This includes the case
of a very weak first order phase transition (too weak, that is, to
satisfy the sphaleron bound), a second-order phase transition and
an analytic cross-over (where there is, strictly speaking, no phase 
transition at all). I consider a two doublet Higgs model (and the
special case which is the minimal supersymmmetric standard model),
and calculate the baryon production due to a CP odd term
\cite{TZandMSTV} in the
one loop effective action ${\cal L}_{eff} =(g^2/16 \pi^2)  \chi F \tilde{F}
= \dot{\chi} B$, where $F$ and $\tilde{F}$
are the $SU(2)$ field strength tensor and
its dual, and the latter form follows from the anomaly equation
and the assumption of homogeneity. The time dependent parameter
is given by \cite{TZandMSTV} 
\begin{equation}
\dot{\chi}= {7\zeta_3}
\left (\frac{m_t}{\pi T}\right)^2 \frac{v_2^2}{v_1^2+v_2^2}\dot{\theta} 
\label{eq: tz} 
\end{equation}
where $v_1$ and $v_2$ are the ratios of the two vevs (the first of which is
assumed to couple to the top quarks), $\theta$ is the CP odd relative 
angle between them, and $\zeta_3 \approx 1.2$.
Performing a simple local thermal equilibrium calculation in the presence
of this term (which is simply a potential for baryon number) 
one obtains the baryon to entropy ratio
\begin{equation}
\frac{B}{s} = -\frac{45 c_n}{2\pi^2 g_*} 
\left(\frac{H}{T}\right)_{ freeze}
\bigg(T\frac{ d\chi }{dT}
\bigg)_{ freeze}
\label{eq: B/s}
\end{equation}
where $c_n \approx 0.44$, $g_*$ is the number of relativistic degrees
of freedom and the subscript means that the quantities are evaluated
at freeze-out of the sphaleron processes.

In order that (\ref{eq: B/s}) be in the range compatible with
nucleosynthesis we therefore require the expansion rate 
at freeze-out to satisfy
\begin{equation}
\left(
\frac{H}{T}\right)_{\rm freeze} 
\simeq (2-12) \times 10^{-11}\, g_*\,
\frac{1}{|(Td\chi/dT)_{\rm freeze}|}.
\label{eq: expansionbound}
\end{equation}

\section{ An alternative cosmology}

The previous section clearly motivates consideration of the
possibility that the expansion rate at the electroweak scale
is considerably greater than usually assumed. One simple way
in which this can come about is if there was some component
of the energy density {\it scaling faster than radiation}
which dominated prior to nucleosynthesis (but red-shifts away
sufficiently by that time). One does not need to look far
beyond the usual framework of early universe cosmology to 
find such a component, for consider a real scalar field $\phi$
with potential $V(\phi)$ for which the equation of motion 
(of the zero mode) is
\begin{eqnarray}
\ddot{\phi} + 3 H \dot{\phi} + V'(\phi)= 0
\label{eq: eompotl}
\end{eqnarray}
Defining $\zeta(t) = \frac{ V(\phi)}{\rho(\phi)}$ where
$\rho(t)=\frac{1}{2} \dot{\phi}^2 + V(\phi)$ is the total
energy density of the scalar field, one gets on integration
that
\begin{eqnarray}
\rho(t) = \rho(t_o) e^{-\int_{t_o}^{t} {6}({1 - \zeta(t)}) H(t) dt}
= \rho(t_o) e^{-\int_{a_o}^{a} {6}({1 - \zeta(a)}) \frac {da}{a}}
\label{eq: scaling}
\end{eqnarray}
Attention is usually focussed on the case $\zeta \rightarrow 1$ 
since it gives inflation with $\rho \approx const $. It is the other
limit, of kinetic energy domination, when $\zeta \rightarrow 0$ and
$\rho \propto 1/a^6$ which is of interest here. A phase of the universe
dominated by such a mode I term {\it kination}.

A very simple way in which such an epoch could follow inflation is if
the inflaton potential becomes asymtyotically  exponential with
$V(\phi) = V_o e^{-\lambda \phi / M_P}$ where $M_P=2.4 \times 10^{18}$GeV 
is the reduced Planck mass. Such potentials can in fact arise quite
generically in theories involving compactified dimensions, 
such as superstring and supergravity theory.  For sufficiently large
$\lambda$ the field will run into a mode which scales as $1/a^6$. 
How then is the Universe reheated? It was pointed in \cite{spokoiny}
that in a model of this sort the energy density in particles created
by the inflationary expansion would come to dominate at some later
stage. Typically at the end of inflation, when the expansion rate is
$H_i$ and the energy density in the inflaton $\rho_i$,
the energy density created in radiation is $\rho_{radn} \sim H_i^4$,
and $\frac{\rho_{radn}}{\rho_i} \sim \frac{\rho_i}{M_P^4} << 1$.
The Universe after inflation is filled with radiation (which
rapidly thermalizes), but dominated until a lower temperature
$\sim H_i^2/M_P$ by the kinetic energy scaling as $1/a^6$. 
During such a phase we have
\begin{equation}
H^2= 
\frac{1}{3M_P^2}\frac{\rho_e}{2}\left[
\left(\frac{a_e}{a}\right)^6 + 
f(a)\left(\frac{a_e}{a}\right)^4
\right]\,,
\label{eq: einsteinb}
\end{equation}
where $a_{e}$ is the scale factor when the density
in the mode becomes equal to that in radiation and
$\rho_e$ is the total energy density at that time. 
The factor $f(a)=[g_*(a_e)/g_*(a)]^{1/3}$ 
describes the effect of decouplings. Nucleosynthesis
provides a lower bound on 
$a_e$, which can easily be converted to an upper bound 
on the expansion rate at freeze-out: 
\begin{equation}
\big( \frac{H}{T} \big)_{freeze} \leq 2 \times 10^{-11}
\left(\frac{T_{\rm freeze}}{100\hbox{\rm GeV}}\right)^2
\label{nucleosynthesis-upperbound}
\end{equation}
Allowing this bound to be saturated (i.e. taking the mode
to dominate as long as is consistent with nucleosynthesis)
then gives the requirement 
\begin{equation}
\left\vert T\frac{d\chi}{dT}\right\vert_{\rm freeze} 
\geq  g_*
\left(\frac{100 GeV}{T_{\rm freeze}}\right) ^2 
\label{eq: CP violation bound}
\end{equation}

A simple examination of the two Higgs doublet model shows that 
in a typical part of the parameter space, one has 
$T\frac{d\theta}{dT} \approx \frac{1}{\tan \theta} 
\frac{T}{v}\frac{dv}{dT}$, where $v$ is one of the vevs.
Using the perturbative effective potential for the minimal 
standard model(MSM), one finds that $\frac{d \phi}{dT} \sim 50$
in a narrow temperature range below $T_c$ where the vev changes by 
about $50\%$. This will typically be where the sphaleron processes
freeze out. Examining the data from lattice studies of the MSM
\cite{kajetalcrossover} this sort of behaviour appears to survive 
in the regime of analytic cross-over - 
physical quantities change continuously, but `sharply' over a range
of a few GeV (with $T_f$ which can be as large as $\sim 250$GeV). 
Thus the constraint (\ref{eq: CP violation bound}) may indeed 
be satisfied in parts of the parameter space of models such as
the two Higgs model with a CP odd source term like (\ref{eq: tz}).

I conclude by returning to the analogy with nucleosynthesis. 
I began by comparing the calculation here to fitting the
Helium abundance predicted by nucleosynthesis by varying the
number of relativistic degrees of freedom. To go further
and emulate nucleosynthesis, we need other probes of
the expansion rate. What might be our deuterium?
An interesting possibility is the relic density
of any dark matter particle which freezes out prior
to nucleosynthesis, the sensitive dependence of which on
cosmology has been previously discussed in the literature
\cite{barrow}. A more indirect root is through
observables consequences at other epochs of alternative
cosmologies prior to nucleosynthesis e.g. the existence of
an exponential field of the type required in the model 
outlined has effects on structure formation in
the Universe, which should be discernible
in future observations of the cosmic microwave background \cite{pgfmj}.

\end{document}